\documentclass[twocolumn,10pt]{article}

\usepackage[round,numbers,sort&compress]{natbib}
\makeatletter
\renewcommand\@biblabel[1]{#1.}
\makeatother

\usepackage{changebar}
\usepackage{times}
\usepackage{graphicx}
\usepackage[english]{babel}

\nochangebars

\title{A master relation defines the nonlinear viscoelasticity of single fibroblasts.}

\author{
Pablo Fern\'andez, Pramod A. Pullarkat, and Albrecht Ott.\\
\\
\\
\normalsize{Physikalisches Institut, Universit\"at Bayreuth}\\
\normalsize{D-95440 Bayreuth, Germany}
\\
\\
}

\begin{document} 

\maketitle 

\fbox{
\begin{minipage}[b]{68mm}
Accepted for publication in Biophysical Journal

Corresponding author: Pablo Fern\'andez

E-mail: pablo@ep1.uni-bayreuth.de


\end{minipage}
}

\section*{ABSTRACT}
{\bf Cell mechanical functions like locomotion, contraction and division are controlled by the cytoskeleton, a dynamic biopolymer network whose mechanical properties remain poorly understood. 
We perform single-cell uniaxial stretching experiments on 3T3 fibroblasts. By superimposing small amplitude oscillations on a mechanically prestressed cell, we find a transition from linear viscoelastic behavior to power-law stress stiffening. Data from different cells over several stress decades can be uniquely scaled to obtain a master-relation between the viscoelastic moduli and the average force. 
Remarkably, this relation holds independently of deformation history, adhesion biochemistry, and intensity of active contraction. In particular, it is irrelevant whether force is actively generated by the cell or externally imposed by stretching. 
We propose that the master-relation reflects the mechanical behavior of the force bearing actin cytoskeleton, in agreement with stress stiffening known 
from semiflexible filament networks.}
\vspace{-2mm}
\section*{INTRODUCTION}

Mechanical forces are essential for biological systems and their interaction with the environment. Tissues can generate forces, but forces also influence tissue development, as in embryogenesis, bone growth or scar tissue formation \citep{fungbook}. In vitro, single eucaryotic cells develop tension spontaneously on a substrate \citep{upto10}, and respond to extracellular elasticity \citep{mechanosensing}. 
Cell mechanics is controlled by the cytoskeleton \citep{alberts,howard}, 
a highly dynamic biopolymer network far from thermodynamic equilibrium. Its main components are actin, tubulin and intermediate filaments. Different types of crosslinking proteins join these filaments into a myriad of geometries. Particularly interesting for cell mechanical properties are actin filaments, which can generate active forces through interaction with myosin motor proteins. 

In the last years, much progress has been made in 
the study of reconstituted biopolymer systems with a known and limited number of components. The rheological behavior of such biopolymer networks has been experimentally studied and explained by simple models \citep{nonlineargels,gardel1,mackintosh1,activefluidization,slowactive}. 
Relating the so gained framework to living cells, however, is a daunting task. Unlike passive networks, the actin cytoskeleton actively generates forces. Further, because of cytoskeleton restructuring or changes in motor activity, the mechanical behavior evolves in time. This evolution is controlled by signaling cascades, which in turn are influenced by mechanical perturbations. 

Probing the mechanical properties of the actin network at the micrometer scale has become possible thanks to newly developed microrheological techniques  \citep{mechanotrans,prestress1,trepat2004,glassy1}. 
However, cell heterogeneity complicates the extrapolation of local properties to the whole cell scale. Many-cell experiments, such as those on tissue or cell-populated gels \citep{modeltissue,fung,cellstotissues}, are efficient and reproducible, but they average over heterogeneous sets of cells interconnected by extracellular matrix. This hampers an interpretation in single cell terms. The quantitative study of the mechanical properties of entire, single cells is thus an essential step towards a global understanding of cell mechanics. 

\begin{figure}[t] \begin{center} \includegraphics*[width=6 cm]{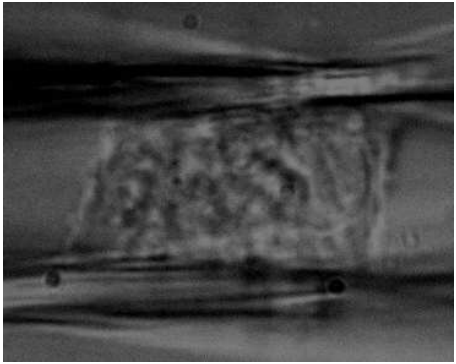} \caption{\small Image of an actively contracting fibroblast between two fibronectin coated microplates.  The distance between plates is $\simeq 5 \mu$m.  } \label{cellfoto} \end{center} \end{figure}

We report uniaxial stretching experiments on single 3T3 fibroblasts, suspended between two parallel, biochemically functionalized walls (see Fig$.\,$\ref{cellfoto}).
This microplate rheometer allows us to clearly distinguish active behavior from the passive mechanical response to applied forces or strains \citep{thoumine}.

\vspace{-1cm}

\section*{MATERIALS AND METHODS}

\subsection*{Experimental setup} 
\begin{figure}[t] \begin{center} \vspace{-0.1 cm} \includegraphics*[angle=0, width=8cm]{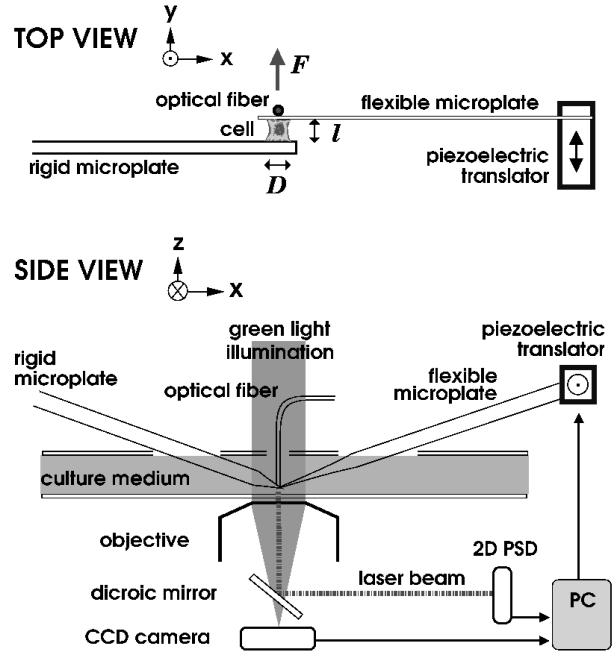} \vspace{-0.1 cm} \caption{\small Schematic of the micromanipulation set-up. A fibroblast is held between two coated microplates. The deformation of the flexible microplate gives the force $F$ acting on the cell. The position of the laser beam emerging from the optical fiber which is in contact with the tip of the flexible plate is detected using a position sensitive detector (PSD). A personal computer reads the signal from the detector and controls the piezoelectric translator.  } \label{cellpusetup} \vspace{-0.4 cm} \end{center} \end{figure}

We describe the main features of our cell-rheometer, schematically shown in Fig$.\,$\ref{cellpusetup}. This is an improved version of the original, home built micromanipulation setup previously described \citep{thoumine}. 3T3 Fibroblasts \citep{3T31,3T32} are held between two parallel, coated glass microplates. One of them is rigid and provides a reference point. The other microplate is thin and narrow and therefore flexible, with an effective stiffness $\sim100$ nN/$\mu$m, which is of the order of the elastic modulus of a typical fibroblast. Its bending provides the force acting on the cell in the {\it y}-axis (see Fig$.\,$\ref{cellpusetup}). The flexible microplate is translated by means of a piezoelectric actuator P-841.40 (Physik Instrumente, Karlsruhe, Germany) with a resolution of 1 nm. An optical fiber in contact with the flexible microplate couples microplate bending to translation of the emergent laser beam. The optical fiber is etched with hydrofluoric acid to a diameter of 6 $\mu$m, so that its stiffness is far lower than that of the flexible plate and does not interfere with the force measurement. The cell is illuminated with green light and observed with an Axiovert 135 microscope (Zeiss, Oberkochen, Germany). A dichroic mirror separates the green illumination light from the He-Ne laser beam, which reaches a S-1880 two dimensional position sensitive detector (Hamamatsu photonics, Japan) through one of the microscope ports. A personal computer reads the signal from the position detector, controls the piezoelectric actuator and calculates the normal force $F$ and the cell length $\ell$. The precision in the measurement of the fiber position is about 100 nm. In most cases a high resolution in time is not needed and averaging can improve the resolution to 30 nm. By controlling the piezoelectric translator, a feedback loop can impose user defined force or length histories. The response time of the piezoelectric translator limits dynamic measurements to frequencies below 30 Hz. Due to the large length of the flexible microplate, its tip deflects by less than 6 arc min during an experiment. Hence, the experimental geometry can be described as two parallel walls, which can be separated by a translation in the perpendicular direction. 

\textit{Temperature Control.} 
The temperature of the cell-culture medium inside the chamber is controlled by means of two ITO-coated glass slides, one on the bottom of the chamber, the other one above, with holes for the microplates and the optical fiber to pass through. In order to avoid convection in the medium, these slides are kept at different temperatures, imposing a temperature gradient pointing upwards with a magnitude of $\sim$ 1 $^{\circ}$C/cm.

\textit{Microplates.} 
The glass microplates used for the experiment are obtained by pulling glass strips (Vitrocom, NJ) as described previously \citep{thoumine}, using a modified P-97 Flaming/Brown micropipette puller (Sutter Instruments, Novato, CA). Microplates are then cleaned and coated with fibronectin from bovine plasma (Sigma-Aldrich) or 3-aminopropyl triethoxysilane (Sigma-Aldrich) and glutaraldehyde (Fluka Chemie, Buchs, Switzerland). 

\textit{Cell culture.}
Experiments where the cells stick to fibronectin-coated plates are performed in ISCOVE medium, with 25 mM HEPES buffer and 10\% fetal bovine serum (FBS). Experiments with a glutaraldehyde coating start in pure saline solution (HBSS) to avoid inactivation of the coating by proteins and amino-acids. After attachment of the cell to both plates, FBS is added to a final concentration of 2\%. All cell culture reagents are from Gibco (Invitrogen, Carlsbad, CA). 3T3 fibroblasts are obtained from the German Collection of Microorganisms and Cell Cultures (DSMZ, Braunschweig, Germany) \citep{DSMZ}.

\section*{RESULTS}
\subsection*{Active response at constant cell-length}
\begin{figure}[!!h] \begin{center} \includegraphics [angle=0, width=8cm] {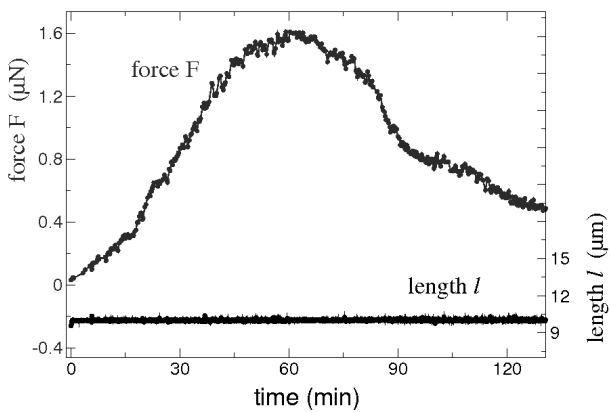} \caption{\small Force as a function of time at constant cell extension, recorded immediately after incorporation of the fibroblast into the rheometer. No significant cell shape alterations are seen throughout.  The experiment is performed using fibronectin mediated adhesion.  } \label{active} \end{center} \end{figure}

We first perform experiments to characterize the response of fibroblasts to their presence in the rheometer (see Fig$.\,$\ref{cellfoto}). To stimulate contractility, we use high serum concentrations of 10\%, and fibronectin mediated adhesion using coated microplates. 
Fibronectin binding to integrins is known to trigger the formation of focal complexes, which connect the extracellular matrix to the actin cytoskeleton \citep{integrins_mechanotrans,choquet}. To minimize the mechanical perturbation to the cell, we keep the cell length $\ell$, given by the distance between the rheometer microplates constant and measure the force $F$. A typical response is shown in Fig$.\,$\ref{active}. After contact with the fibronectin coatings, most cells generate contractile tension, the force $F$ reaches values up to 1 $\mu$N and eventually decays to zero. The force relaxation is an adaptation to the constant length condition, since active contraction can be induced again by a sudden change in cell length $\ell$. The behavior is reproducible only in its broad, qualitative features. The force and time scales are strongly cell-dependent. 

\begin{figure*}[t]
\begin{center}
\includegraphics*[angle=0, width=17 cm]{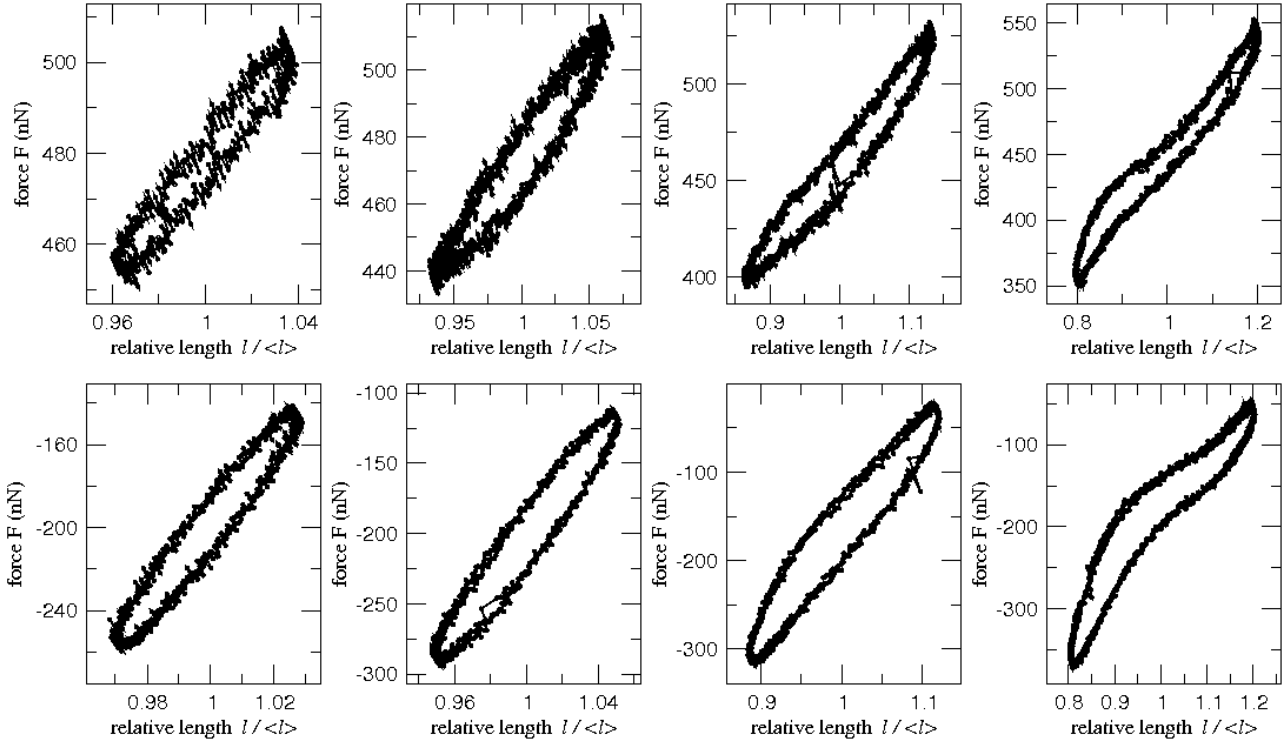}
\caption{\small
Lissajoux figures for different amplitudes. 
{\it Top)} Typical response at forces above crossover. Force $F$ as a function of relative cell length $\ell/\langle\ell\rangle$ for strain amplitudes 3.5\%, 6\%, 13\%, and 20\%. The oscillation frequency is 0.2 Hz.
{\it Bottom)} Typical response at compressive forces. Force $F$ as a function of relative cell length $\ell/\langle\ell\rangle$ for strain amplitudes 3\%, 5\%, 12\%, and 20\%. The oscillation frequency is 0.2 Hz.
}
\label{overtones}
\end{center}
\end{figure*}

\subsection*{Dynamic viscoelastic moduli}
In order to probe cell rheological properties during complex adaptive processes, we superimpose sinusoidal length oscillations at small amplitudes and high frequencies. These are chosen so that the corresponding maximum rate of change of force is at least two orders of magnitude above the values typically seen in active contraction. 

Cells have both an elastic and a viscous response to imposed length oscillations \citep{fungbook,onaran}. As the viscous response depends on deformation rate instead of absolute deformation, it causes a phase-shift between length and force.
The cell length is imposed as \[
\ell = \langle\ell\rangle + \Delta_\ell \sin(\omega t),
\]where $\langle \ell \rangle$ is the average length over an oscillation period, and $\Delta_\ell$ is the amplitude of the superimposed length-oscillations.  
In the linear regime, force is then given by 
\[ F = \langle F\rangle + \Delta_F^\prime \sin(\omega t) + \Delta_F^{\prime\prime} \cos(\omega t). \]We now introduce the initial contact area $A_0 = \pi\,(D/2)^2$, where $D$ is the apparent cell diameter. The amplitudes are related by 
\begin{eqnarray*}
\frac{\Delta_F^\prime} {A_0} = \Theta^\prime \, \frac{\Delta_\ell}{\langle\ell\rangle} \;
&\quad\mbox{and} \quad&
\frac{\Delta_F^{\prime\prime} }{A_0} = \Theta^{\prime \prime} \, \frac{\Delta_\ell}{\langle\ell\rangle},
\end{eqnarray*}
where $\Theta^\prime$ and $\Theta^{\prime\prime}$ are differential viscoelastic stretching moduli. The storage modulus $\Theta^\prime$ reflects the purely elastic (non-dissipative) part of the cell reaction, and the loss modulus $\Theta^{\prime\prime}$ the viscous (dissipative) contribution. In order to compare cells of different sizes and with biological gels, we use formal engineering stress units for the moduli. Accordingly, stress is taken as $\sigma=F/A_0$.

\cbstart
The differential stretch moduli $\Theta^\prime$ and $\Theta^{\prime\prime}$ should not be confused with material parameters like the Young's modulus. The spatial distribution of force bearing structure inside the cell is unknown. Rather than introducing ad hoc hypotheses, such as assuming a uniform material, we simply treat the cell as a mechanical black box. The unconventional symbol $\Theta$ for the moduli intends to emphasize their experiment-specific nature. Further, these moduli characterize the response of the material to small perturbations around a situation which may be far away from the resting state. Indeed, we show below that it is suitable to study $\Theta^\prime$ and $\Theta^{\prime\prime}$ as a function of the average force $\langle F\rangle$. 
A similar approach has been successfully used in stretching experiments on whole tissues, such as skin or muscle \citep{fungbook}. Equivalent procedures have recently been applied to biopolymer gels under shear deformations, where the differential shear moduli are measured as a function of the average stress or strain \citep{nonlineargels, gardel1}. 
\cbend

Instead of the loss and storage moduli $\Theta^\prime$ and $\Theta^{\prime\prime}$, it will be more convenient to regard the absolute modulus $\vert\Theta\vert$ and the loss angle $\delta$, defined as 
\begin{eqnarray*}
\vert\Theta\vert=\sqrt{\Theta^{\prime\;2}+\Theta^{\prime\prime\;2}}	\\
\delta = \arctan\left(\frac{\Theta^{\prime\prime}}{\Theta^\prime}\right)
\end{eqnarray*}

\begin{figure}[b!] \begin{center} \vspace{-0.0cm} \includegraphics*[angle=0, width=8 cm]{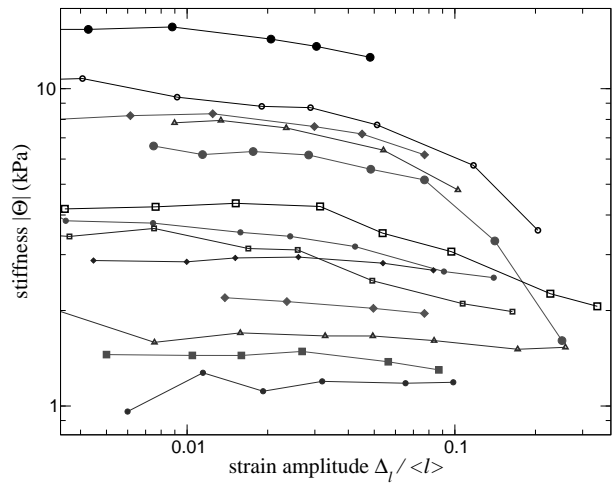} \vspace{-0.0 cm} \caption{\small Amplitude sweep. Stiffness $\vert\Theta\vert$ as a function of the strain amplitude $\Delta_\ell / \langle\ell\rangle$ for an arbitrary selection of cells. Each curve is a different experiment. All frequencies are 0.2 Hz.} \label{ASweep} \vspace{0.0 cm} \end{center} \end{figure}

Fig$.\,$\ref{ASweep} shows the dependence of the stiffness $\vert\Theta\vert$ as a function of the strain amplitude $\Delta_\ell / \langle\ell\rangle$. In general, at strain amplitudes in the range 0.02--0.06, the moduli do not change by more than 20\%. 
Moreover, no significant distortion of the response is seen below relative deformations of 0.1, as shown in Fig$.\,$\ref{overtones}. 
This holds irrespective of the frequency in the range 0.1--1 Hz. Thus, in subsequent experiments the amplitude is kept small, $\Delta_\ell = 0.5\;\mu$m, which corresponds for all cell length values to 0.02--0.06 strain amplitudes.
\cbstart
More amplitude sweeps, including the amplitude dependence of the loss angle $\delta$, can be found as supplementary material. 

\begin{figure*}[t!] \begin{center} \includegraphics*[angle=0, width=13.5cm]{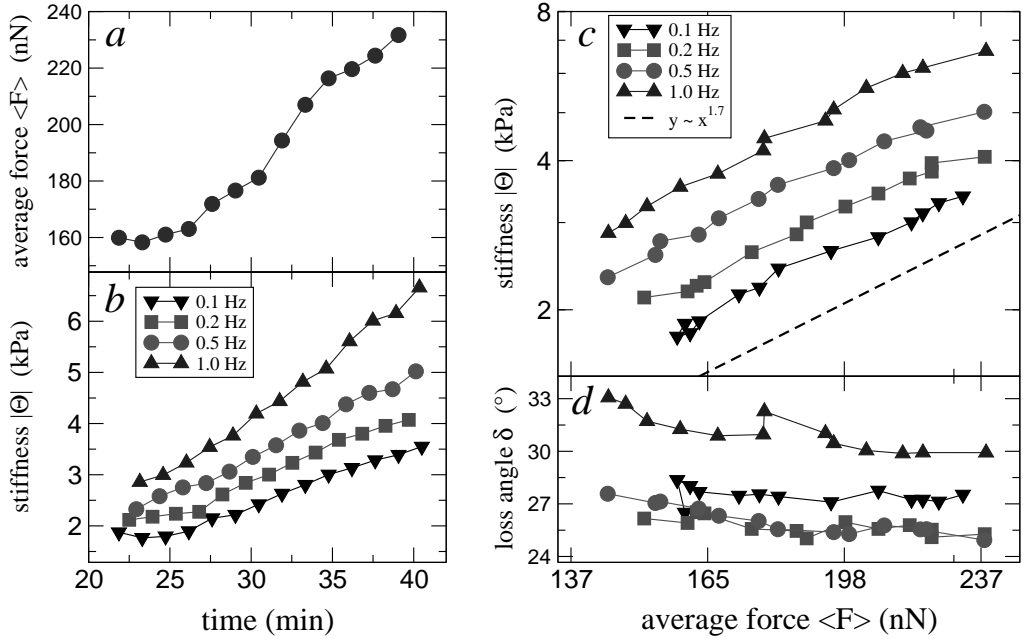} \caption{\small Stress stiffening at constant length. The average cell length is kept constant throughout, $\langle\ell\rangle = 9 \mu$m.  Sinusoidal oscillations are superimposed with a strain amplitude $\Delta_\ell / \langle\ell\rangle = 0.03$. The frequency of the oscillations is cyclically changed from 0.1 to 1.0 Hz. {\it a)} The average force $\langle F\rangle$ is seen to increase in time. {\it b)} The modulus $\vert\Theta\vert$ increases in time for all frequencies. {\it c)} Stiffness $\vert\Theta\vert$ as a function of average force $\langle F\rangle$, for different frequencies. The line shows a power-law function $y \sim x^{1.7}$. {\it d)} Loss angle $\delta$ as a function of average force $\langle F\rangle$, for different frequencies.  } \label{Anfang} \end{center} \end{figure*}

\begin{figure}\vspace{-0.5cm}
\begin{center}
\includegraphics*[angle=0, width=8 cm]{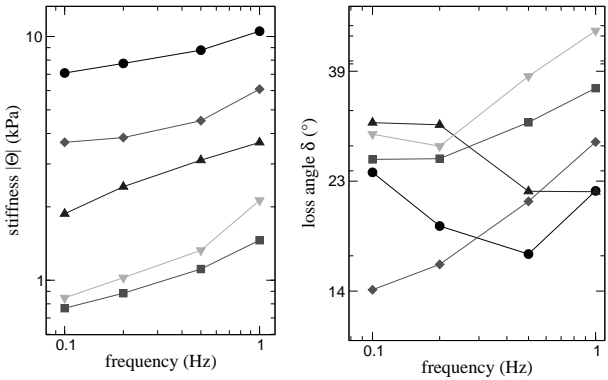}
\caption{\small
Frequency sweep (preliminary data). Modulus $\vert\Theta\vert$ and loss angle $\delta$ as a function of frequency. Each curve is obtained using a different cell. 
}
\label{fsweep}
\end{center}\vspace{-1.6cm}
\end{figure}

For completeness, in Fig$.\,$\ref{fsweep} we show preliminary results on the frequency dependence of the modulus $\vert\Theta\vert$ and the loss angle $\delta$, in the range 0.1--1 Hz. We are currently modifying the setup in order to explore a wider frequency range, as at present we are limited to just one frequency decade. Keeping this in mind, our results are consistent with the literature on cell mechanics. On the average, $\vert\Theta\vert$ increases weakly with the frequency, whereas $\delta$ is approximately constant. The modulus $\vert\Theta\vert$ increases with the frequency roughly as a power-law with exponents in the range 0.1--0.3. Such a frequency dependence, the signature of a flat, broad continuous spectrum of relaxation times, has been known in soft tissues for a long time \citep{fungbook} and has recently been observed at the micrometer scale \citep{glassy1,glassy2}. Accordingly, recent experiments at the whole-cell scale performed with a setup similar to ours have revealed power-law creep functions \citep{singlecreep}. 
\cbend

\pagebreak

\subsection*{Stress stiffening at constant cell length}

During the initial phase of force development after contact with the fibronectin-coated microplates, the cell sweeps force-space at a constant length. We superimpose sinusoidal oscillations to the constant average length, in order to probe the temporal evolution of the moduli $\Theta^\prime$ and $\Theta^{\prime\prime}$. The frequency of the oscillations is cyclically changed in the range 0.1 -- 1.0 Hz. As shown in Fig$.\,$\ref{Anfang}, as the average force increases with contractile activity of the cell, so does the modulus $\vert\Theta\vert$. Figs$.\,$\ref{Anfang} {\it c, d} show the dependence of the response parameters $\vert\Theta\vert$ and $\delta$ on the average force $\langle F\rangle$ for different frequencies. This is an example of active stress stiffening, since it takes place at an average constant length.

\subsection*{A master-relation characterizes stress stiffening}

\begin{figure}[!h] \begin{center} \includegraphics [angle=0, width=8 cm] {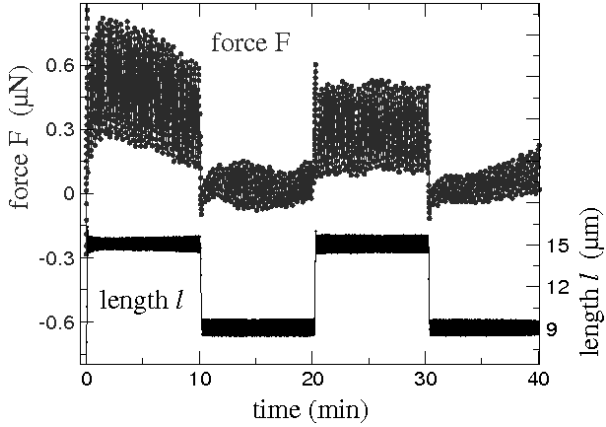} \caption{\small The force response as result of imposed length changes. We step-strain the cell by about 50\% at a rate of 1.5 $\mu$m/s, and apply length oscillations at an amplitude of 0.5 $\mu$m and a frequency of 0.2 Hz. The experiment was performed using fibronectin mediated adhesion.  } \label{step-stretch} \end{center} \end{figure}

Not all fibroblasts generate such high forces as in the experiment shown in Fig$.\,$\ref{Anfang}. In order to span a larger range of average force $\langle F\rangle$ and length $\langle\ell\rangle$, we step-stretch the cell and then keep the average length constant, superimposing oscillations to record the moduli $\Theta^\prime$ and $\Theta^{\prime\prime}$. The procedure is periodically repeated, as shown in Fig$.\,$\ref{step-stretch}. As a reaction to a sudden change in length 
a force relaxation always occurs, usually followed by active contraction. As the average force $\langle F\rangle$ evolves at a fixed length $\langle\ell\rangle$, the viscoelastic moduli are continuously recorded. We also perform step experiments controlling the average force $\langle F\rangle$. In this way, it is possible to span large areas in the $\langle\ell\rangle$ -- $\langle F\rangle$ diagram. 

\begin{figure}[t!] \begin{center} \includegraphics [angle=0, width=8.5cm] {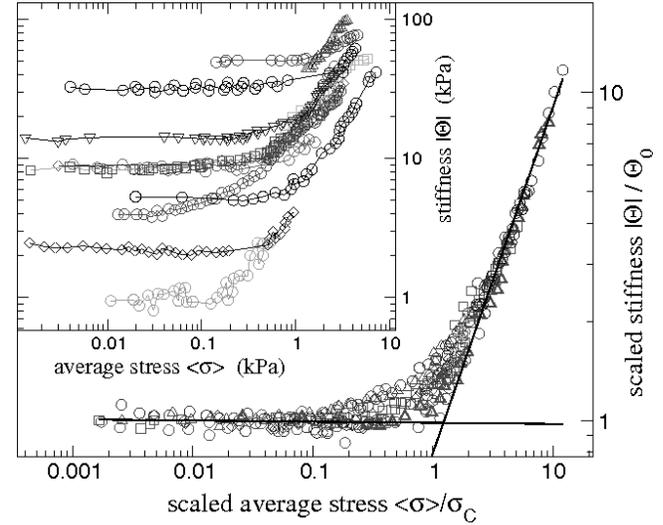} \caption{\small The inset shows the elastic modulus $\vert\Theta\vert$ as a function of average stress for 13 cells, measured using length steps plus oscillation experiments such as in Fig$.\,$\ref{step-stretch}. The main plot shows the data scaled using 2 factors, which gives an exponent $X\simeq 1.0$. All experiments are performed at 26$^{\circ}$C and using fibronectin mediated adhesion.} \label{master} \end{center} \end{figure}

We see that both viscoelastic moduli depend only on the average force, and are independent of the average length. 
The dependence of the loss angle $\delta = \arctan(\Theta^{\prime\prime}/\Theta^\prime)$ on the average force $\langle F\rangle$ is erratic and weak, at most decreasing about 20\% in the whole force range. As a function of the individual cell, it is within the range 10-30$^\circ$. 
The absolute modulus 
$\vert\Theta\vert=(\Theta^\prime{\hspace{0.5mm}}^2+\Theta^{\prime\prime}{\hspace{0.5mm}}^2)^{1/2}$ remains constant at low forces, in a 1--30 kPa range depending on the individual cell.
Above a cell-dependent crossover force, we observe stress stiffening: $\vert\Theta\vert$ increases as a function of the average force $\langle F\rangle$. This dependence of $\vert\Theta\vert$ on the average force can be well approximated by a power-law, as shown in Fig$.\,$\ref{master} \textit{(inset)}.
More than one stress decade above crossover, most cells deform significantly and begin to detach or yield. 

A collapse of all data to a single master-relation can be achieved by introducing cell-dependent scaling factors, the zero force stiffness $\Theta_0$ and the crossover stress $\sigma_C$. On the average, 
\[
\vert\Theta\vert = \left\{ \begin{array}{r@{\quad\mbox{for}\quad}l}
                   
\Theta_0 \quad\quad\quad\:\,\:
& \langle\sigma\rangle < \sigma_C \\

\Theta_0  \left(\frac{\textstyle\langle\sigma\rangle}
{\textstyle\sigma_C}\right)^{\textstyle \gamma} 
& \langle\sigma\rangle >\sigma_C

\end{array} \right. 
\]

The exponent $\gamma$ is independent of the scaling factors. At $26^\circ$C, 0.2 Hz, and 5\% deformation amplitude, it is approximately 1, as shown by the collapsed data in Fig$.\,$\ref{master}. The scaling factors are roughly related by $\Theta_0 \propto \;\sigma_C^{\: 1.3}$. Thus, an approximate collapse can be reached with a single parameter.

This master-relation is consistently found in all cells strong enough to reach average stresses above $\sim 0.1\:\Theta_0$. This reproducibility shows that oscillatory measurements are effective in probing cell mechanical properties independently of an underlying slow active behavior.

\subsection*{Ramp experiments with superimposed oscillations}
\begin{figure}[t] \vspace{2mm}
\begin{center} \includegraphics [angle=0, width=8cm] {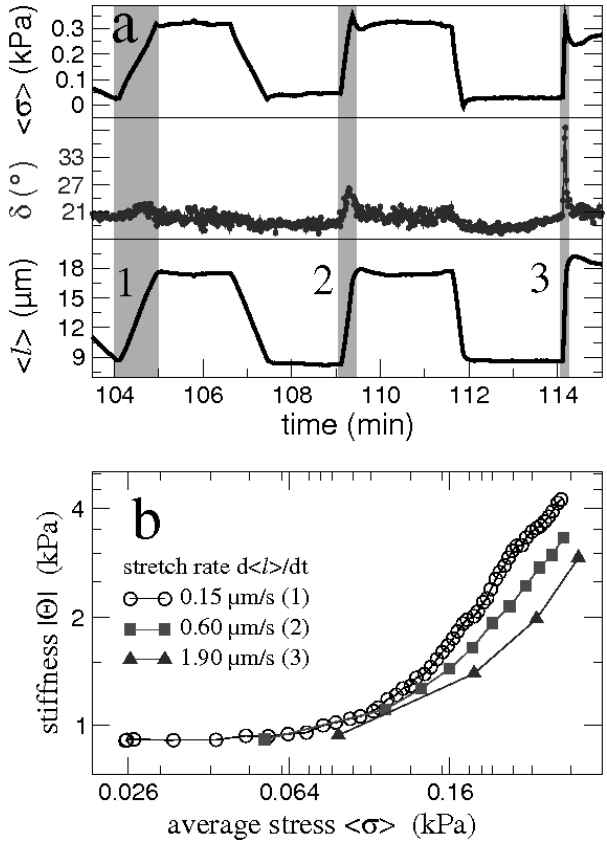} \caption{\small {\it a)} Ramp experiment with glutaraldehyde coating. Average stress {\it (top)}, loss angle {\it (middle)} and average length {\it (bottom)} as a function of time. Oscillations at 1 Hz, 5\% amplitude are superposed throughout, also during the ramps {\it (shaded areas)}. The phase difference $\delta$ increases with the deformation rate.  \newline {\it b)} Stiffness as a function of average stress. The relationship between $\vert\Theta\vert$ and $\langle \sigma\rangle$ depends on the deformation rate. The curves correspond to the shaded areas 1, 2, 3 in Fig$.\,$\ref{stretching} {\it a}.  } \label{stretching} \end{center} \vspace{-5mm}
\end{figure}

The master-relation holds at a constant cell length while an underlying active contraction occurs. In this case, the rate of change of the average stress is set by the cell. As a means of assessing the validity of the master-relation while the average cell length changes, we increase $\langle\ell\rangle$ at a constant rate in the range 0.1 -- 2 $\mu$m/s, and simultaneously superimpose small oscillations at an amplitude $\Delta_\ell \simeq 0.5\:\mu{\rm m}$ and a frequency of 1 Hz, to measure the dynamic moduli $\Theta^\prime$ and $\Theta^{\prime\prime}$. 
As shown in Fig$.\,$\ref{stretching} {\it a}, by stretching the cell, a change in average stress is induced. The average stress $\langle\sigma\rangle$ depends roughly linearly on the average length $\langle\ell\rangle$ throughout a ramp. Remarkably, stress stiffening of the dynamic moduli is simultaneously observed. The master-relation between $\vert\Theta\vert$ and $\langle \sigma\rangle$ is seen to remain valid at low deformation rates. 
These experiments show that the particular way of sweeping force space is not relevant, since the $\vert\Theta\vert(\langle\sigma\rangle)$ function is qualitatively the same to that found in active contraction experiments. 
Only at rates higher than a cell-dependent value in the vicinity of 100 nm/s, $\vert\Theta\vert(\langle\sigma\rangle)$ falls below the master-relation, and the cell becomes more fluid as evidenced by an increase in the loss angle $\delta$ (see Fig$.\,$\ref{stretching}, {\it a} and {\it b}). In general, the master-relation cannot be deduced from the relationship between the average values $\langle\sigma\rangle$ and $\langle\ell\rangle$, since these are linearly related in a ramp experiment. This absence of stiffening agrees well with results from ramp experiments on whole tissues \citep{fung, fungbook} and fibroblast populated collagen gels \citep{modeltissue}.

\vspace{-1mm}
\subsection*{Cytoskeleton perturbation using drugs}
In order to distinguish between different cytoskeletal subcomponents, experiments are performed in presence of drugs, which disrupt specific filament types. Only cells able to sustain tensions above $\sigma_C$ are subsequently treated with drugs. 
We observe sudden detachment of the cells from glutaraldehyde-coated microplates with the actin depolymerization inducer Latrunculin-A at 1 $\mu$g/ml \citep{Latrunculins}. The same effect is seen with the myosin heavy chain ATPase inhibitor 2,3-butanedionemonoxime at 2 mg/ml \citep{BDM}, as well as with the myosin light chain kinase inhibitor ML-7 at 100 $\mu$M \citep{ML7}. 
Since transmembrane proteins are covalently bound to the glutaraldehyde coating, they must rip off from the membrane during drug-induced cell detachment. Thus, the cell membrane alone is not able to hold transmembrane proteins under significant tension. An internal structure must bear the load under normal conditions. After disruption of either actin or myosin activity, this structure cannot sustain strong forces anymore. 
At 4-fold lower concentrations, Latrunculin-A and ML-7 reduce $\vert\Theta\vert$ up to a factor of 5, while $\langle\sigma\rangle$ goes to zero. No significant effect is seen with the microtubule-disrupting drug Nocodazol at 10 $\mu$g/ml \citep{nocodazol}. 
Taken together, these results show that the actomyosin system bears the tension, without any other significant force-bearing structure in parallel. 

\vspace{-1mm}
\subsection*{Stress stiffening with glutaraldehyde coatings}
In order to reduce active responses, we use glutaraldehyde-aminosilane coated walls, where accessible membrane proteins are covalently and non-specifically bound through imine-groups. Further, the serum concentration is reduced to $2\%$. In this way, the biochemical conditions are changed, but the experimental cell geometry remains the same. Active responses are indeed found to be weaker in these conditions. Compared to fibronectin mediated adhesion, the stiffness values are scaled down by about a factor of three. However, the master-relation between $\vert\Theta\vert$ and $\langle\sigma\rangle$ can still be observed.

\section*{DISCUSSION}

The master-relation is a general feature, which reproducibly occurs in our experiments. 
It holds during active contraction and adaptive responses, such as those seen in Figs$.\,$\ref{active}, \ref{Anfang} and \ref{step-stretch}, as well as in ramp experiments with superimposed oscillations, as in Fig$.\,$\ref{stretching} {\it b}. 
It is observed in different biochemical environments, which change the intensity of active responses. 
From this generality of the master-relation, we conclude that it reflects a fundamental property of the cell force-bearing elements.

\subsection*{Stress stiffening in actin networks}

Our cytoskeleton perturbation experiments point towards the actin network as the main component responsible for stress stiffening. Actin filaments, with a persistence length of 15 $\mu$m \citep{howard, persistence}, behave as semiflexible filaments in vivo, where typical filament lengths are about 1 $\mu$m. 
It has recently become apparent that crosslinked as well as entangled networks of semiflexible filaments show characteristic nonlinear mechanical behavior. 
In vitro prestressed biopolymer gels under shear deformation, including crosslinked actin networks, generally show a transition from a linear regime to
power-law strain stiffening \citep{nonlineargels,gardel1,janmey1,xu,gelatin1}.
The connection to our results has to be established with care. In vitro gels are passive minimal systems, generally studied under shear deformations. Instead, the living cell is a complex entity where a multitude of crosslinking proteins are available, many of them of a dynamic nature. Cytoskeletal restructuring might take place within an oscillation period. Keeping this in mind, we still find the qualitative similarity between the stiffness-force relations very suggestive. In scruin-crosslinked actin networks \citep{gardel1}, the ratio between the crossover stress and the zero force modulus is of the order of $10^{-1}$, and the stress-stiffening exponent is in the range 1--1.5, in remarkable agreement with our results. In experiments on filamin-crosslinked actin gels \citep{nonlineargels}, where the storage modulus $G^\prime$ is actually studied as a function of strain, the crossover strain is also of this order.

\vspace{-1mm}
\subsection*{Intrinsic stress stiffening}

The master-relation connects the viscoelastic moduli to the average force independently of cell length. Rather than strain stiffening, fibroblasts exhibit intrinsic \textit{stress} stiffening. To reconcile this to our interpretation of the master-relation, 
we postulate that the internal strain of the gel, i.e., the stress, uncouples from the cell length $\ell$ as a result of active contraction, via rearrangements of the connection points of the network. 

As shown by ramp experiments with superimposed oscillations, the master-relation holds even during an externally 
imposed length increase, if the extension rate is sufficiently slow. The magnitude of the deformations which can then be imposed (more than 50\%), without the cell yielding, suggests that addition of material takes place. 
As the loss angle $\delta$ does not 
change under such conditions, we speculate that the cytoskeleton 
grows without major structural changes. The cytoskeleton is certainly capable of such restructuring, as exemplified by the growth under compression of the actin tail of Listeria \citep{pramana,yanns}, towed growth in neurites \citep{axon_elongation_mech} or stress fibers losing material while contracting \citep{solacontra_stressfi}. The maximal growth speed 
from our ramp experiments would be around 6 $\mu$m/min, in good agreement 
with the growth of the actin tails of the bacteria Listeria \citep{pramana}. At higher deformation rates the cell starts to 
fluidize, which can be interpreted by an increased fraction of ruptured bonds. 


\subsection*{Master-relation in the context of cell mechanics}
\vspace{-0.5mm}
Several features of the master-relation can be observed in experiments performed at very different length scales, corroborating our interpretation as a general feature of the force-bearing elements in the cell. 
This shows the master-relation to be general, independent of the type of mechanical experiment or cell geometry. 
Uniaxial stretching experiments performed on skin, myosin fibers, tendons \citep{cellstotissues,fungbook,fung} and on fibroblast-populated collagen gels \citep{modeltissue} show proportionality between dynamic stiffness and force. These results can be seen as a particular case of power-law stress stiffening, for an exponent $\gamma=1$. 

Our results are also similar to stress stiffening in skeletal and smooth muscle, where stiffness is proportional to force. The generally accepted explanation is that both stiffness and force are a function of the variable number of actomyosin crossbridges \citep{muscle1}. Although such an explanation is attractive, 
it does not seem to apply to our case. We have measured forces up to $1\;\mu$N, in agreement with total forces exerted by spreading fibroblasts of $\sim2\;\mu$N \citep{upto10,stressloco}. This corresponds to $\sim10^6$ myosins working in parallel, very close to the total amount reported in fibroblasts \citep{totalactin}. Taking crossbridge stiffness as 0.6 pN/nm \citep{howard}, an arrangement of $10^6$ myosins in parallel would be a factor of 100 stiffer than the maximum $\vert\Theta\vert$ we have measured in fibroblasts. 

At the subcellular scale, microrheology experiments performed on adhering cells show proportionality between the shear storage modulus $G^\prime$ of the actin cortex and the force applied by the cell on the substrate \citep{prestress1}. Here, the force was increased by stimulating cell contractility with histamine, or decreased by the relaxing agonist isoproterenol. The result compares well to our observation of stress stiffening at constant length, though no crossover to a linear regime is reported here. Recently, simultaneous increase of both the storage and loss shear moduli $G^\prime$ and $G^{\prime\prime}$ of the actin cortex after stretching epithelial cells has been observed \citep{trepat2004}. Accordingly, we see stress stiffening of both longitudinal moduli $\Theta^\prime$ and $\Theta^{\prime\prime}$ when stretching the cell in a ramp experiment with superimposed oscillations. Thus, the fact that the master-relation holds regardless of the way force-space is explored agrees with these microrheological studies taken together. 

A similar example of length scale invariance is given by the agreement between microrheological studies in the frequency domain \citep{glassy1,glassy2} and whole-cell experiments in the time domain  \citep{singlecreep}. 
Fabry et al \citep{glassy1,glassy2}, studying the shear moduli $G^\prime$ and $G^{\prime\prime}$ of the actin cortex as a function of frequency, could collapse data from cells in different conditions onto master curves. The response in the frequency domain as revealed by these microrheological procedures was subsequently found to agree with the power-law creep function of single cells \citep{singlecreep}. 
Hence, scale invariance and self similarity may well be general properties of cell mechanics in the force, as well as in the time-frequency domain.

\pagebreak
\subsection*{Possible stiffening mechanisms}

Aiming at describing crosslinked biopolymer networks, theoretical frameworks have been developed recently. Networks are modeled as a 2-dimensional random arrangement of filaments, characterized by a bending and a stretching modulus. Temperature effects have been addressed by including an entropic stretching modulus \citep{semiflex,mackintosh1}. At a non-zero temperature, thermal energy is stored in bending fluctuations, so that the distance between crosslinks is smaller than the actual contour length of the filament. Separating these crosslinks by stretching reduces the amplitude of these shape fluctuations, which gives rise to an increased effective stretching modulus. In general, the mechanical response of 2-dimensional random networks depends strongly on the length scales involved, the filament length and the mesh size $\xi$ (the average distance between filaments). 
For long filament lengths or small mesh sizes, the deformation of the network is affine, i.e., the macroscopic mechanical response is essentially given by that of single filaments \citep{mackintosh1}. 
The nonlinear elasticity of crosslinked biopolymer networks has been explained in terms of random networks \citep{gardel1,nonlineargels}, as an affine stretching-dominated regime where the macroscopic response is given by single-filament entropic stretching. This response is highly nonlinear, since thermal fluctuations are suppressed as the filament ends are separated close to full contour length. At high forces, the stiffness-force relation becomes a power-law with exponent $3/2$ \citep{gardel1}.  

An alternative explanation for stiffening in crosslinked random networks has been recently presented by Onck et al \citep{entropiciscrap}. Here, stiffening arises as a transition from bending to stretching. Thermal effects are mostly irrelevant; increasing the temperature from 0 to 300 K increases the crossover strain, but does not affect the stiffening regime. Also the transition to stiffening arises 
in spite of a certain degree of non-affinity.

\begin{figure}[t!]
\begin{center}
\includegraphics [angle=0, width=8cm] {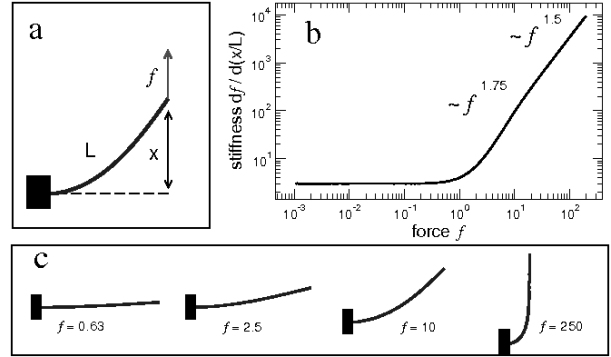}
\caption{\small
{\it a)} Single-filament nonlinear elasticity: filaments are regarded as inextensible, characterized by a contour length $L$ and a bending modulus $\kappa$. One end is clamped, the other one free. The force $F$ bends the filament by an amount $x$. The dimensionless force is defined as $f=F L^2/\kappa$.
\newline
{\it b)} Differential stiffness $df/d(x/L)$ as a function of the dimensionless force $f$. 
\newline
{\it c)} Euler-Bernoulli \textit{elastica}: filament shapes for different forces. 
}
\label{elastica}
\end {center}\vspace{-4mm}
\end{figure}

Below we point out some remarkable features of single filament bending response which may be relevant for crosslinked filament networks. We discuss the bending response of inextensible filaments, characterized by a length $L$ and a bending modulus $\kappa$. We show that bending itself shows a crossover to power-law stress stiffening with an exponent 1.75, as well as a crossover strain of the order of $10^{-1}$. This can be seen in Fig$.\,$\ref{elastica}. At large forces, the stiffness-force relation becomes a power-law with exponent $3/2$, similar to the entropic stretching response. The details of the calculation are left to the appendix. 
It is noteworthy that many features of the master-relation are similar to the mechanical response of single filaments. The experimental fact that a single parameter is sufficient to obtain the master relation may be captured by both the force and stiffness scales varying as $\kappa/L^2$. 
Interestingly, the magnitude of the experimental crossover stress, when expressed as a strain $\sigma_C/\Theta_0$, is of the order of $10^{-1}$, in good agreement with the theoretical crossover strain. 
As Fig$.\,$\ref{elastica} {\it b} shows, the first force decade above crossover of the theoretical stiffness-force relation is an approximate power-law with exponent 1.7, within the experimentally observed range. 
Further, the magnitude of the force scale is the right one if one assumes a realistic cytoskeleton mesh size of $100$ {\rm nm} \citep{foam,howard,tomography} and an actin bending modulus of $\kappa = 60\;\rm{nN(nm)}^2$ \citep{howard}.
This corresponds to the open-cell foam model proposed by Satcher and Dewey as a general model for the cytoskeleton \citep{foam}, which gives a zero-force stiffness $\Theta_0 \sim 10\; {\rm kPa}$, in good agreement with our measurements and with literature \citep{foam,cellstotissues}. 

These observations suggest that the bending response of filaments may play a role in the nonlinear mechanical properties of crosslinked gels. Therefore, it is worthwhile considering this effect in further modeling of such systems. With the present set of experiments, we cannot distinguish between the different mechanisms proposed for stiffening. Further experimental work and theoretical modeling will be required to elucidate the relative contributions of these effects for physiologically relevant values of mesh sizes and filament lengths.

\subsection*{Summary, conclusion and outlook}

By performing single fibroblast mechanical 
measurements we have revealed a master-relation relating stress and 
dynamical cell stiffness. This relation is obtained by a simple scaling of data from different cells. 
For the measurement 
we use the fact that cells are active: as a reaction to a sudden 
perturbation the cell sweeps a range of mechanical stresses. 
We probe the cell elastic properties during this active response. 
In spite of this underlying complex behavior which may involve multiple biochemical pathways, the master relation is surprisingly 
simple and reproducible. 
If the average force is externally changed, by slowly stretching the cell in a ramp experiment while superimposing small-amplitude oscillations, the master-relation between $\langle F\rangle$ and the stiffness $\vert\Theta\vert$ is seen again. Thus, a distinction between active and passive stress is artificial -- regardless of deformation history, the response to small perturbations is always given by the average force. 
In view of this generality, and the remarkable agreement with the mechanic behavior of biopolymer networks, we interpret the master-relation as revealing the nonlinear response of the actin network. 

These experiments provide a framework where simplified biomimetic systems can be connected to living cell mechanics. 
Here, the active nature of the cytoskeleton is naturally integrated into the phenomenology. Along these lines, we are currently extending our frequency and amplitude range, to undertake a more detailed analysis of the master-relation.

As a final remark, fibroblasts reach the stress-stiffening regime 
naturally by active contraction and we expect the crossover force to be within the 
physiological force-range. Thus stress stiffening may play a role in 
vivo. For example, cells are known to sense the extracellular elasticity and exert forces along lines of maximum effective stiffness, 
a process which has been modeled within linearized elasticity \citep{mechanosensing}. 
In the future, power-law stress stiffening could be included in the modeling of cell organization by mechanosensing.

\vspace{0.7 cm}

\begin{footnotesize}
We thank A. Bausch, G. Massiera, H. Brand, L. Mahadevan, O. Thoumine, F. J\"ulicher and R. Everaers for helpful discussions, and B. Wehrle-Haller for his generous gift of GFP-actin fibroblasts. Assistance from A. Hanold in taking care of the cell culture is gratefully acknowledged. We also thank Y. Marcy and A. Micoulet for work on the setup. Finally, we are very grateful to the reviewers for constructive criticism and helpful suggestions, and to the editor for his interest in the manuscript. This work has been funded by the Universit\"at Bayreuth and the Institut Curie (Paris). 
Some of the results of this work were presented in the PHYNECS workshop in Peyresq, France, September 2002, and in the DPG Fr\"uhjahrtagung, Regensburg, Germany, March 2004.
\end{footnotesize}

\section*{Appendix}

\subsection*{Bending response of an inextensible filament}
In what follows we show that a transition to power law stiffening is already present in the nonlinearity of bending deformations. We analyze the mechanical response of inextensible filaments characterized by a bending modulus $\kappa$ and a contour length $L$. We assume that the relationship between bending moment and curvature stays linear, so that any nonlinearity is of geometrical nature. The filament shape then corresponds to Euler-Bernoulli \textit{elastica} \citep{love}. Our boundary conditions are one filament end free and the other one clamped perpendicular to the force direction, as illustrated in Fig$.\,$\ref{elastica} {\it a}. These conditions are the minimum requirements for bending. 
At small strains $x/L$, the mechanical equilibrium equation can be linearized and the zero-force stiffness analytically calculated as $dF/dx=3\kappa/L^3$. 

Numerically solving the nonlinear equilibrium equation, we find a relatively sharp crossover to a strain-stiffening regime at strains $x/L>0.3$, as shown in Fig$.\,$\ref{elastica} {\it b}. At large strains, the stiffness-force relation asymptotically becomes a power-law with an exponent 1.5. At high forces, the filament shape tends to a straight line along the force direction, with a sharp kink at the clamped end (see Fig$.\,$\ref{elastica} {\it c}). In this limit, the effect of increasing $F$ can be shown to be a scaling of the filament shape at the clamped end; the endpoint coordinates and curvature radius remain linearly related as they change as a function of $F$. Thus, we can introduce a force-dependent scaling factor $\varepsilon$, which tends to zero as the force is increased to infinity. In terms of this factor, the full contour length $L$ and the projection length $x$ (see Fig$.\,$\ref{elastica} {\it a}) can be written as $L - x \propto \varepsilon$, and the torque balance equation as $F \varepsilon \propto 1 / \varepsilon$. Therefore, $F \propto 1 / (L - x)^2$, which gives power-law stiffening, $dF/dx \propto F^{\:3/2}$.

In general, an entropic contribution to the longitudinal filament stretching modulus is expected due to thermal fluctuations. At low forces, this leads to a thermal longitudinal modulus $k_{T}=90\kappa^2/(k_B TL^4)$ \citep{semiflex}, where $L$ is the contour length, and $k_B T$ the thermal energy. Taking a mesh size of 100 nm, this thermal longitudinal modulus $k_{\,T}$ is about a factor of $10^4$ larger than the low-force mechanical bending stiffness, $k_{\bot}=3\kappa/L^3$. Since $k_\bot/k_{\,T} \propto L$, the mechanical response of a regular network such as the open-cell foam model \citep{foam} remains dominated by bending stiffness for physiological mesh sizes. 
Further, for such mesh sizes the mechanical bending energy already exceeds the thermal energy at strains above $\sim 4\%$, well within the linear regime. Thus, thermal effects are not relevant for the bending response itself.

\bibliography{Fernandez_etal}

\end{document}